\documentclass{kluwer}
\usepackage{graphicx}
\begin{document}
\begin{article}
\def\magm{$^{\rm m}\!\!\!.\,$}
\begin{opening}
\title{Modeling the Dust Shell of V4334 Sgr}
\author{S.~Kimeswenger}
\institute{Institut f\"ur Astrophysik, Leopold--Franzens Universit\"at Innsbruck\\ 
Technikerstrasse 25, A-6020 Innsbruck, Austria}
\author{J.~Koller}
\institute{Department of Physics and Astronomy, MS-108, Rice University\\
6100 Main Street, Houston, TX 77005, U.S.A.}

\begin{abstract}
The central star V4334 Sgr (Sakurai's Nova) of the planetary nebula PN G010.4+04.4 
underwent in 1995-1996 the rare event of a 
very late helium flash.
It is only one of two such events during the era of modern astronomy 
(the second event was 
V605 Aql = Nova Aql 1919). All other prominent
objects of that type  originate from events several thousands of years ago 
(e.g. A30, A78). Hence, only snapshots can be modeled for those objects.
V4334 Sgr allows for the first time a dynamic consideration of the formation 
of the dust shell from the beginning.
We present here a model which is able to describe the complete photometric behavior
of the object, including the fine structure dips of the optical light curve 
during the first two years of the mass loss and the dust formation.
\end{abstract}
\end{opening}

\section{Introduction}
V4334 Sgr, discovered 1996 February 20 (Nakano et al. \citeyear{Na96}),
was first identified as a nova.
The nature of the object has been clarified by Duerbeck \& Benetti (\citeyear{Du96}) 
as the rare event of a 
very late helium flash shell burning (Iben et al. \citeyear{Ib83}; Iben \citeyear{Ib84}).
For a review on this object see Duerbeck et al. (\citeyear{Du00}).
It is only one of two such events during the era of modern astronomy 
(the second event was 
V605 Aql = Nova Aql 1919, reviewed by Clayton \& de Marco \citeyear{Cl97}). 
Older prominent
objects of that type,  were the event occurred several thousands of years ago, are
A30 (Borkowski et al. \citeyear{Bo94}) and A78 (Kimeswenger et al. \citeyear{Ki98}).
The rapid formation of a dust shell, discovered in March 1997 (Kimeswenger et al.  
\citeyear{Ki97}),
started in the end of 1996. Several snapshot models of this object were published 
(Kerber et al. \citeyear{Ke98}, 
Kipper \citeyear{Kip99a}, Kipper \citeyear{Kip99b}, 
Kerber et al.  \citeyear{Ke99}). 
They all suffer from the fact that a snapshot allows only 
the use of an average dust formation rate.
Thus, they are unable to describe the visual photometry and give only crude estimates
for the density distribution and for the mass of the shell.
The model we present here only uses the visual light curve to determine the
variations of the dust formation rate. It applies a complete dust grain size distribution
and transiently heated grains (Koller \& Kimeswenger 
\citeyear{Ko00a}, \citeyear{Ko00b}).
Only the input from the fitting of the visual light curve gives us a complete 
spectral energy distribution (SED) model,
assuming closed spherical geometry.

\section{Input Data and Assumptions}

A set of data over the whole range of wavelengths
is needed to perform a reliable modeling of the SED. 
But the main input is obtained from wavelengths of 0.3 to 15\ $\mu$m.
While there are numerous optical and near infrared (NIR) photometries, there is a lack of
data in the mid infrared range (MIR). 
The data used here for the MIR were mainly obtained by our work group 
using the ISO satellite (Kerber et al. \citeyear{Ke99}).
The optical photometry were taken from Duerbeck et al. (\citeyear{Du97}, 
\citeyear{Du00}) but
corrected for some errors (1997 was shifted by about 60 days in figure 4 in the 
\citeyear{Du00} paper).
The NIR photometries were taken from Kimeswenger et al. 
(\citeyear{Ki97}) and Kamath \& Ashok (\citeyear{Kam99}).\\
We assume spherical geometry of the dust shell and a variation of the dust formation
rate at timescales not faster than 5 days. 
The grain size distribution was assumed to be like those found in the
numerical simulations by Gauger et al. (\citeyear{Ga90}) and 
Patzer et al. (\citeyear{Pa98}) for winds
of C-rich stars.
For computations the numerical code NILFISC (Koller \& Kimeswenger 
\citeyear{Ko00a}, \citeyear{Ko00b}) was used. 
A distance and an interstellar foreground extinction has to be assumed.
We used here, employing a weighted average from the literature, 
the somehow "canonical" values of $D = 3$\,kpc (Kimeswenger 2001, this volume) and
E$_{B-V}$\,=\,0\magm8 (Kerber et al. \citeyear{Ke00}).

\section{The Model}
The stellar luminosity is the main input parameter for the dust shell model. 
As the shell hides the star efficiently since the beginning of 1997, 
we had to derive it indirectly. 
We know the effective temperature in 1996 rather well. Later 
we have lower limits by the excitation of 
the C$_2$ bands. A change of the effective temperature from 
4800 to 6200 K results only in small changes
of the model parameters. We assume a continuation of the temperature 
decline seen in 1996. This
results in 6150 K in February and 5700 K in October 1997. In this 
temperature range  V-I and the 
bolometric correction for giants and supergiants are entirely 
constant (Bessel \citeyear{Be90}).
This allows us to derive the increase of the luminosity and the rise of the total
shell extinction independently. The visual photometry gives 
a linear incline of the luminosity with time.
\begin{figure}
\centerline{\includegraphics[width=4.85cm]{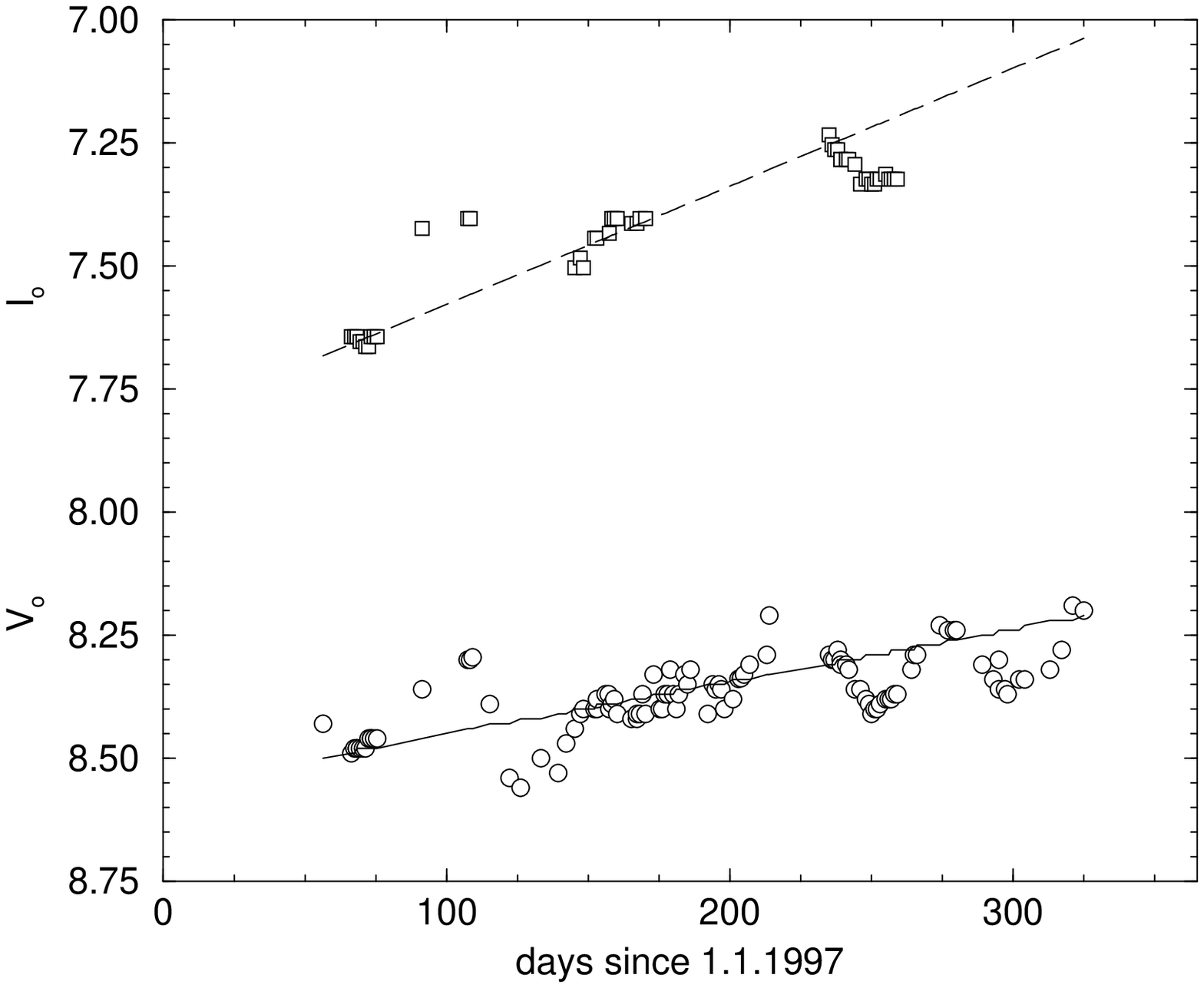}
\phantom{XX}\includegraphics[width=4.85cm]{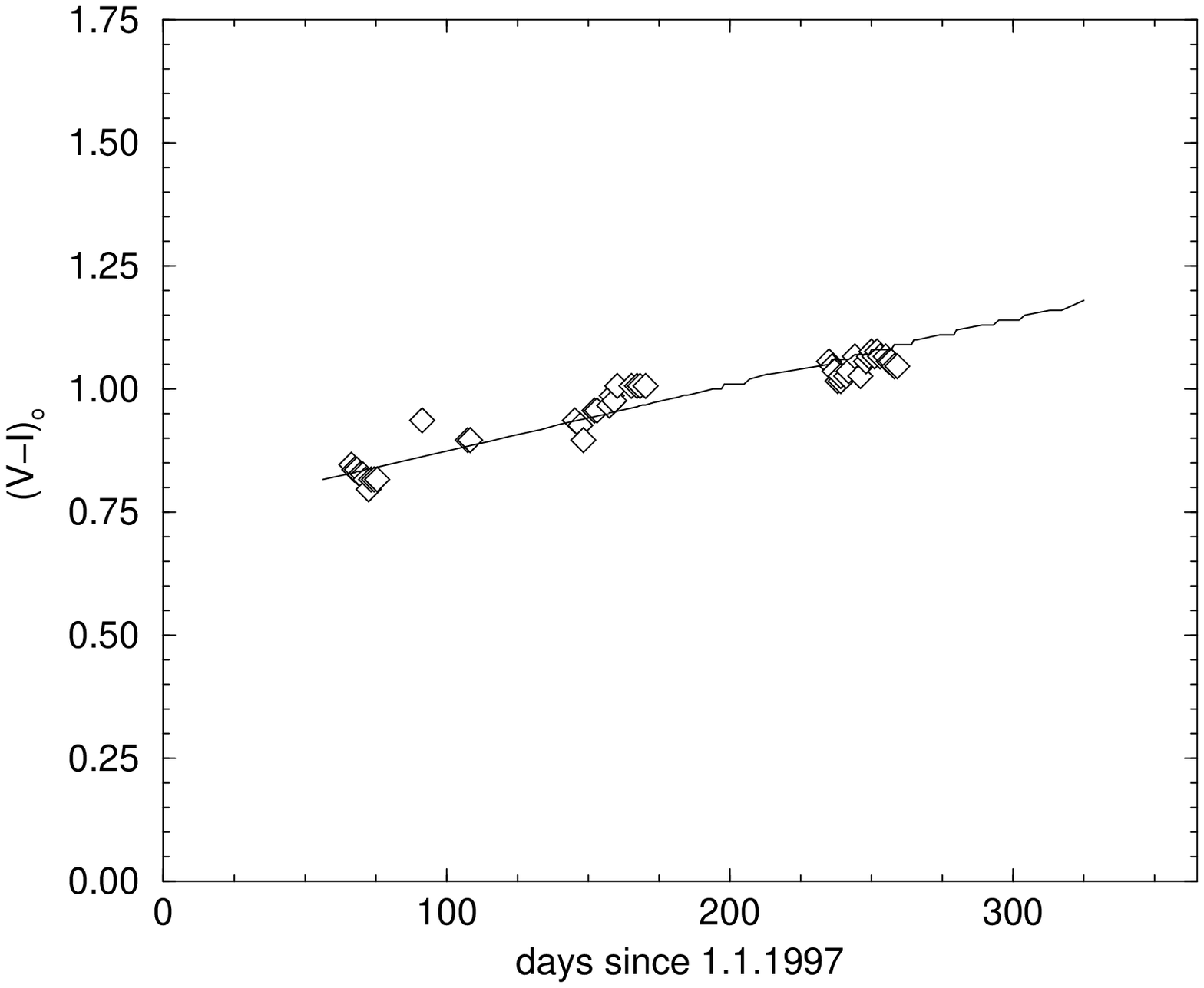}}
\caption{The low frequency evolution of the photometry during 1997.
Left: the I and V band photometry; right: the resulting red color index.
The lines indicate
the relations used to derive the increase of the stellar 
luminosity and of the extinction of the shell. }
\end{figure}
Usually a single expanding shell model is assumed. This model 
is described by 3 free parameters, 
the inner and the outer radius and the density at the inner 
edge. We assume a set of episodic mass loss changes.
The main mass loss episode started in August 1996. 
This is 
defined by the observations. On the one hand there
is  the absence of major extinction features in the late 
1996. On the other hand there is the dust shell
giving the strong IR excess in February 1997. 
The inner radius of the 
condensation (respectively the condensation temperature) and the 
grain size distribution (power law $a^{-2.8}$)
is taken from theoretical models for the formation of carbon 
rich dust in fast stellar winds (Dominik et al. \citeyear{Do93},
Patzer et al. \citeyear{Pa98}, \citeyear{Pa99}). We assume that the conditions
are constant throughout the whole period.
This appears reasonable since the physical and chemical environment 
hardly changed.
Thus, the inner radius is directly bound to the stellar luminosity.
The dust formation occurs in a relatively small region. A further reprocessing of
the dust behind this region is impossible (Dominik et al. 
\citeyear{Do93}) and, therefore, we are fixed to
a $r^{-2}$ density law during one outflow episode. 
If the mass loss rate changes, the shell detaches from the formation 
region and moves outwards without changing its initial velocity. 
A new shell is formed inside. This defines strictly
the outer radius of the shells. The density (respectively the 
mass loss rate as function of time) is now the only "free"
parameter to be derived. As shown before, the visual photometry 
allows us to fit this parameter. 
A sudden outburst of the mass loss rate leads to "triangle shaped 
declines" of the visual light. 
They peak at the end of the episode and then quickly recover 
to the normal value. Those "triangle shaped declines"
are typically for the whole light curve of V4334 Sgr in 1997 
and 1998. They last always about 15 days. This allows us to model
the mass loss rate as a function of time very accurately. In 
1997 major outbursts were on days 110, 235 and 280 and a weak
 but longer one starting on day 155 (45 days long).
\begin{figure}  
\centerline{\includegraphics[width=4.85cm]{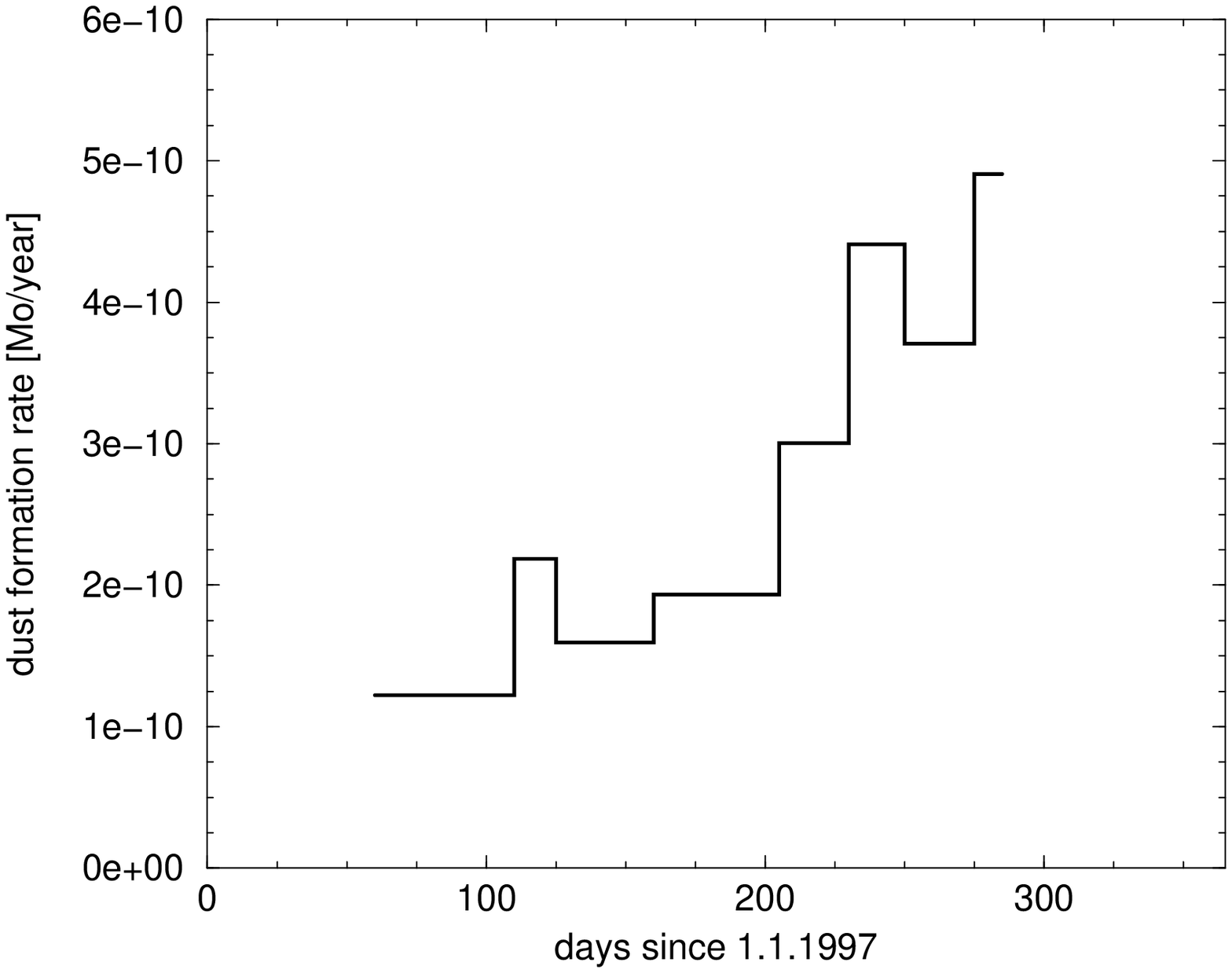}
\phantom{XX}\includegraphics[width=4.55cm]{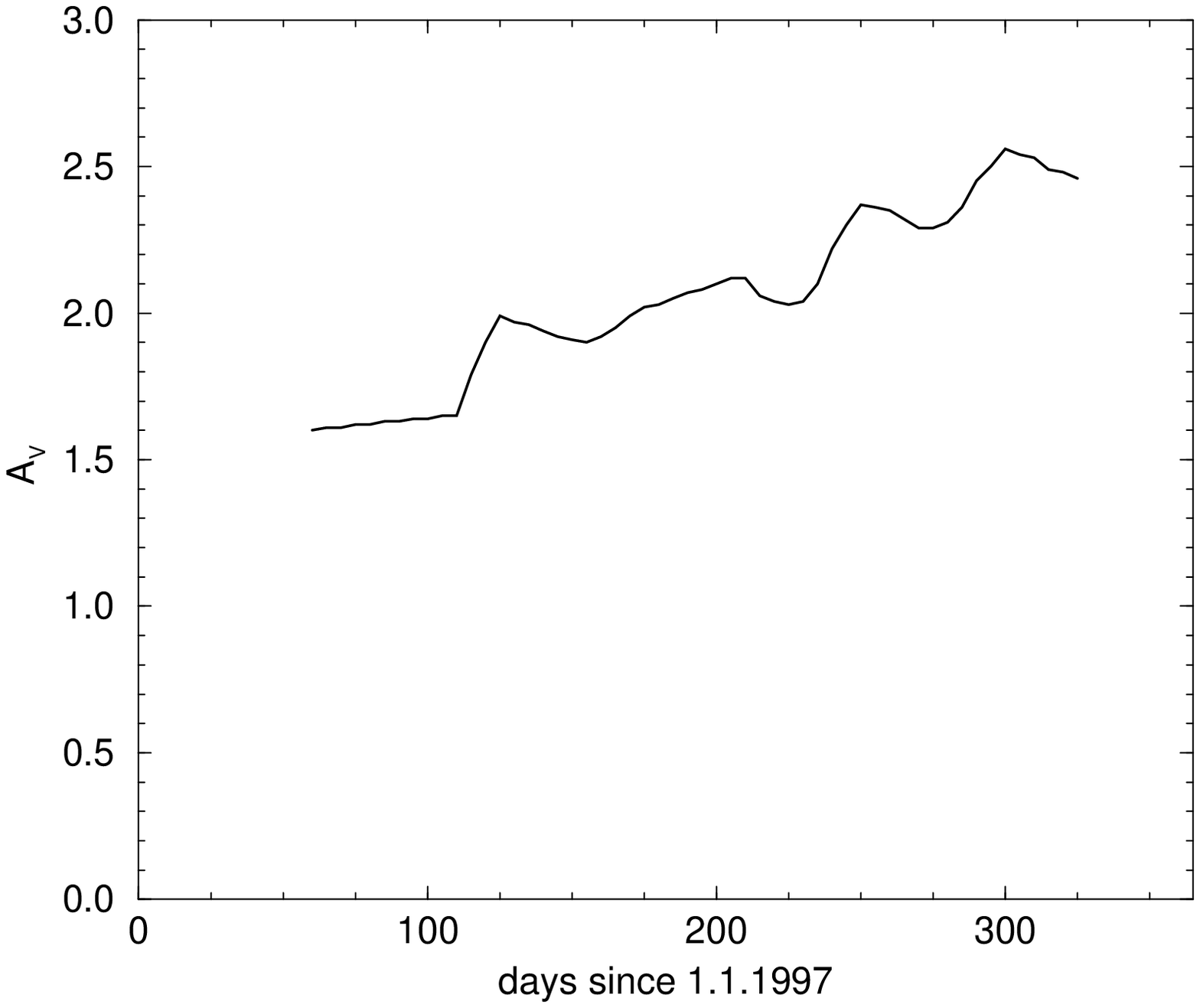}
}
\caption{The dust formation rate (left) resulting from 
fitting the declines in the visual light curve. This gives 
a steady increase of the overall extinction (right) with minor humps.}
\end{figure}
Exactly these model inputs, without any further changes, were used 
to derive complete spectral energy
distributions for 3 epochs where ISO measurements are 
available. The resulting spectral energy 
distributions fit very well to the data. 
\begin{figure}[t]
\centerline{\includegraphics[width=5.3cm]{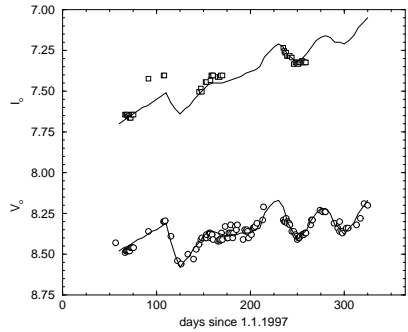}}
\caption{The resulting optical light curve fits very 
well, not only in the V (standard deviation $\sigma$ = 0\magm02), 
but also in the I band ($\sigma$ = 0\magm02).
The few U light curve points, although reproduced completely, are not plotted here. 
Too few points are available to allow
a detailed investigation of the fine structures ($\sigma$ = 0\magm06). }
\end{figure}
\begin{figure}
\centerline{\includegraphics[width=5.5cm]{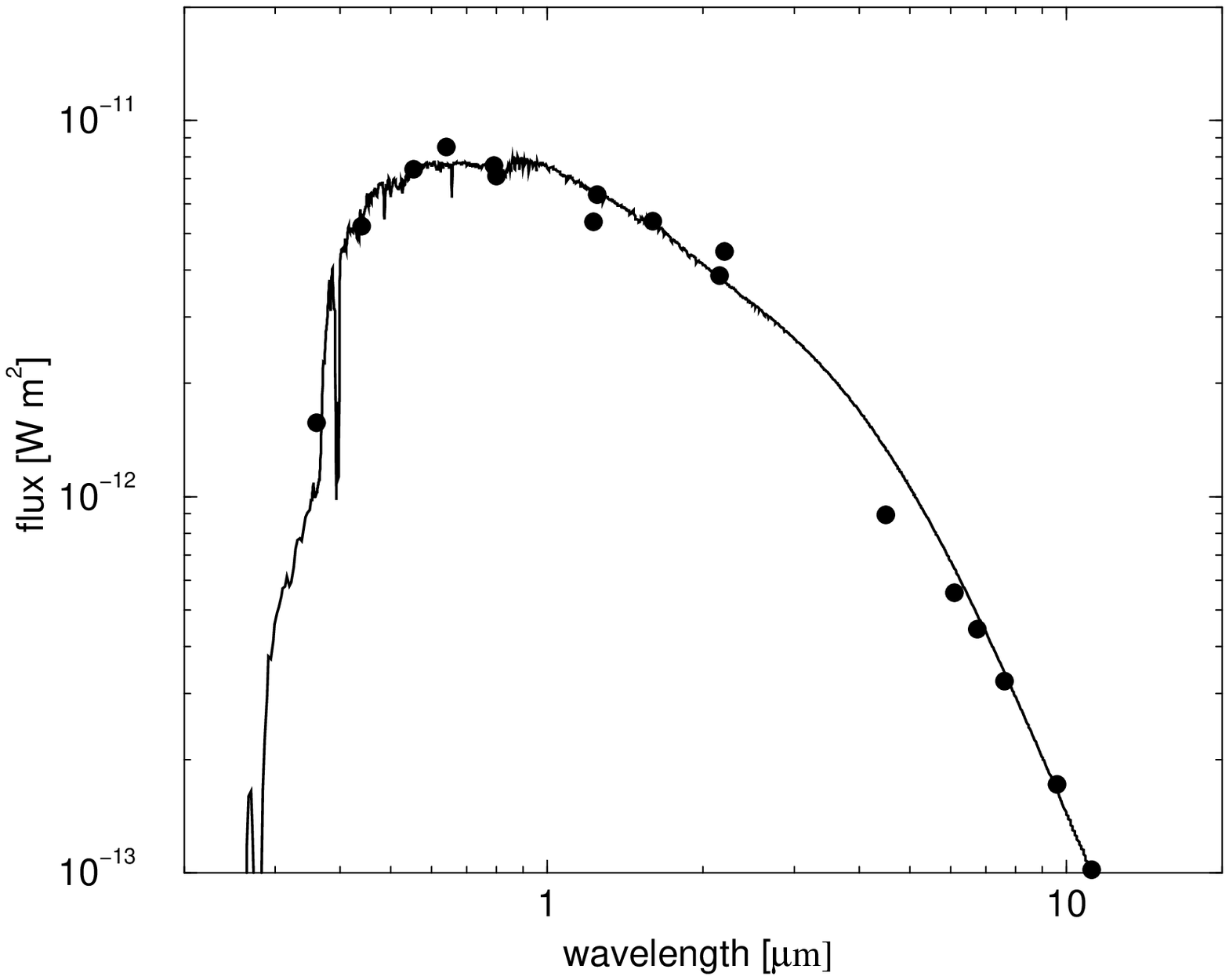}\phantom{XX}
\includegraphics[width=5.5cm]{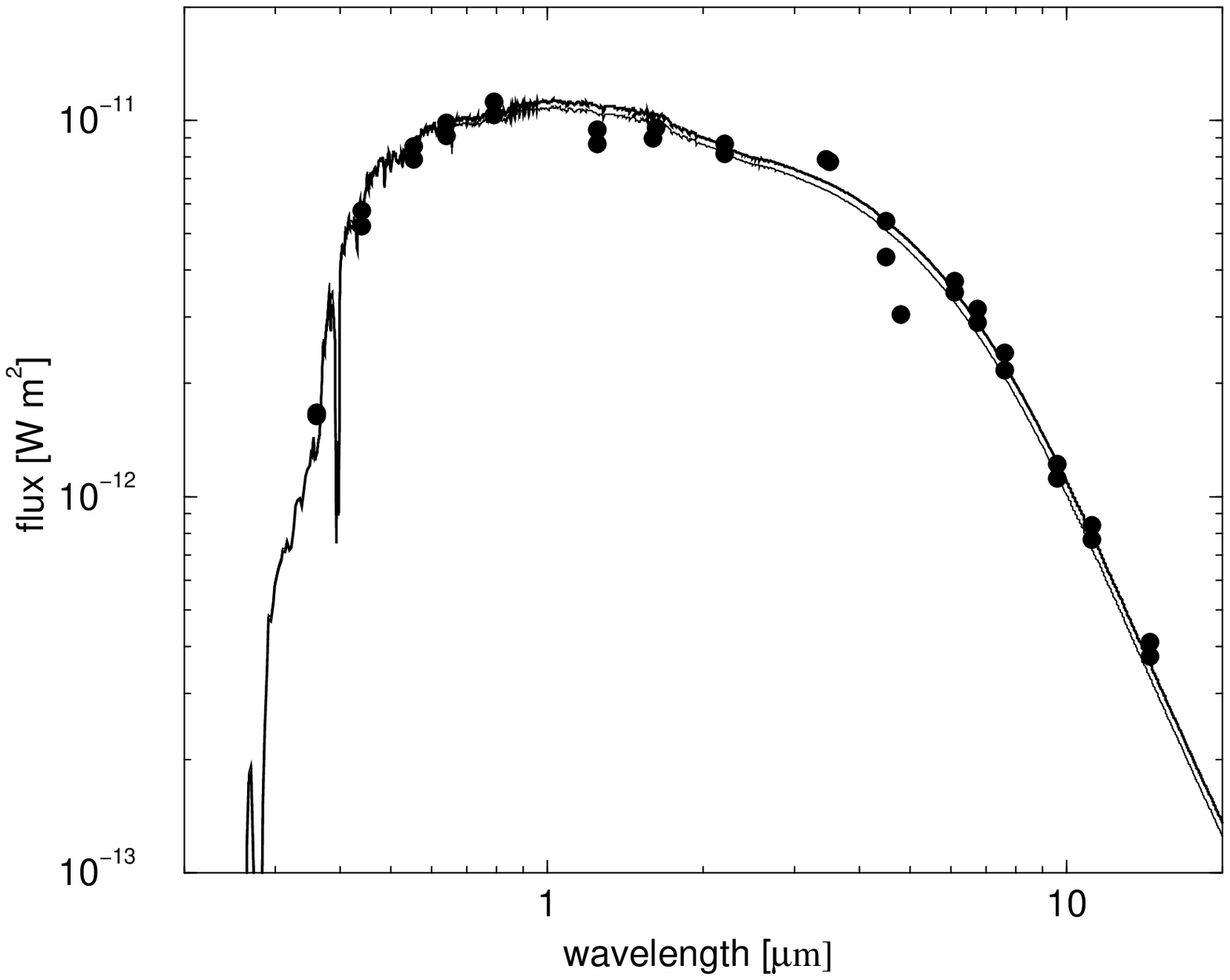}}
\caption{The spectral energy distributions for February 
1997 (left) and September + October 1997 (right)
resulting from our models. The data points are from Duerbeck et al. (2000 - optical), 
Kimeswenger et al. (1997 - NIR), Kamath \& Ashok (1999 - NIR), 
Arkhipova et al. (1998 - 1 to 5 $\mu$m) 
and from ISOCAM (Kerber et al. 1999).
The NIR data in the February 1997 plot by Kamath \& Ashok 
were taken in April and thus are located a little bit too high. The 5 $\mu$m
point of Arkhipova et al. is always below any other reported 
M band photometry (also in summer 1997, for which we computed no model 
due to the lack of ISO data). We think that this 5 $\mu$m point is not reliable.}
\end{figure}
These promising results convinced us that the 
self--consistent access to the problem works.
As we derive the optical density along the line
of sight from the central star to the observer 
only, we can't verify the assumption of
spherical geometry with these results. Tests with cylindrical 
geometries (open at the poles) show that the SED - especially at 
long wavelengths, can't be reproduced. We 
conclude that we have at least a completely closed geometry.
The integrated dust mass evolved rapidly and increased 
by more than a factor of 3 during 1997. The dust formation rate
goes nearly linear with the stellar luminosity.
This implies the consequence that the dust formation rate
is bound to the mass loss rate (respectively the 
dust to gas ratio is more or less constant).  
The outer shell radius increased between February 1997
until October 1997 from $4.5\times 10^{12}$ m to $1.3\times 10^{13}$ m. 
Although the stellar luminosity raised by a factor 
of 2.7, the inner shell radius expanded 
only from $9.0\times 10^{11}$ to $1.23\times 10^{12}$ m. This 
is significantly below the expected 
case for thermodynamic equilibrium for dust grains.
\begin{figure}
\centerline{\includegraphics[width=5.5cm]{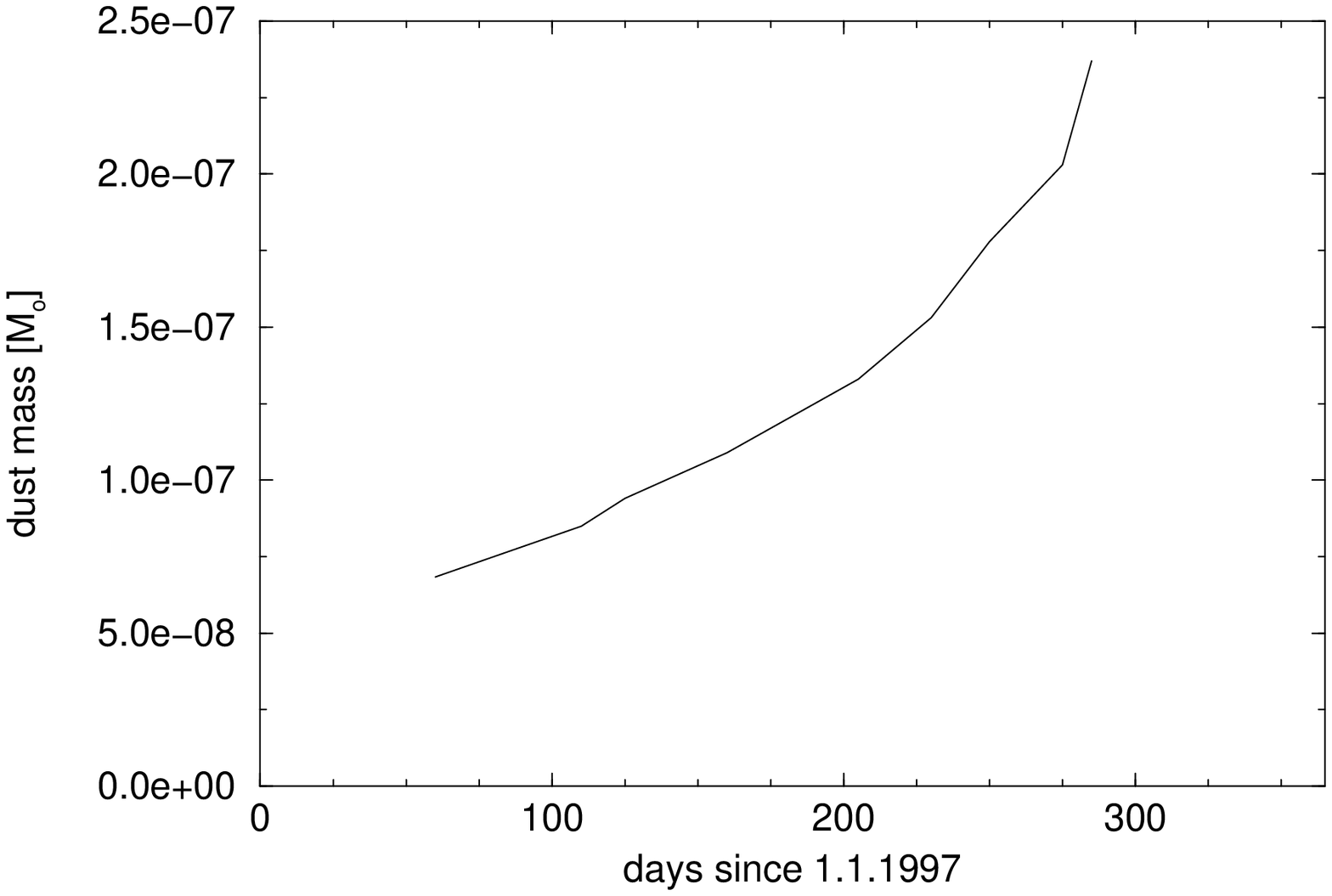}
\phantom{XX}\includegraphics[width=5.5cm]{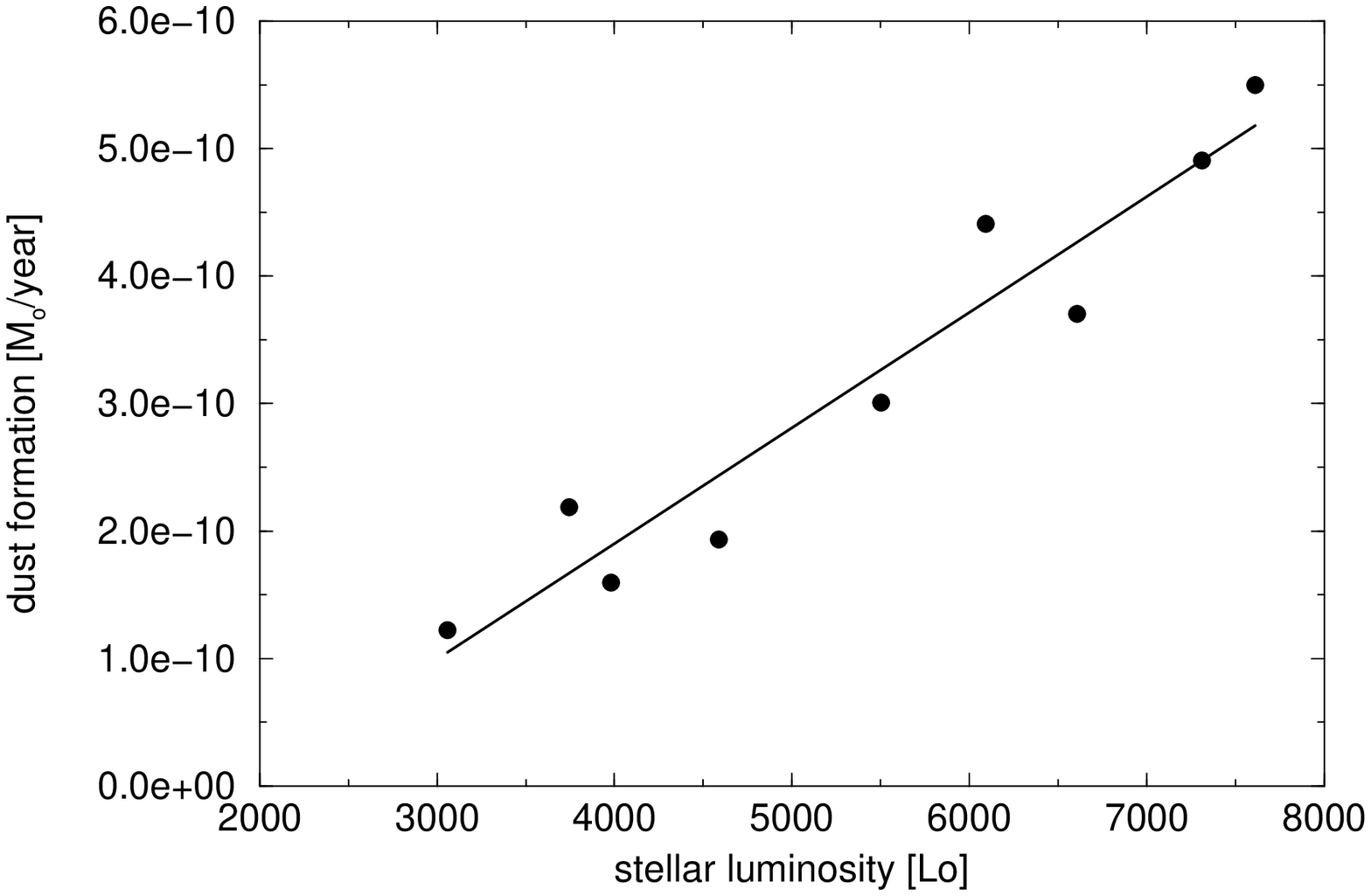}}
\caption{The dust mass (left) and the stellar luminosity (right) increases by more of a 
factor of 3 resp. 2 during 1997.}
\end{figure}
\section{Conclusions}
We showed that a complete dynamic model of the 
early phase of dust formation of this peculiar object is possible
by only applying the optical data to fit the parameters. 
The only free parameter is the 
distance to V4334 Sgr. The resulting model is completely 
self consistent. The dust formation rate increased during 
1997 (at an assumed distance of $D = 3$ kpc) steadily 
from $1.2\times 10^{-10}$  to about $5.0\times 10^{-10}$ M$_\odot$/year
during this first year.  As the dust/gas ratio cannot 
be determined from measurement in this 
system, this also cannot be
used to derive the total mass loss straight forward. 
Assuming a gas to dust ratio of 10 - 30,
 the mass loss would be about 
$10^{-10}$ M$_\odot$/year which is not unusually high.
The extremely carbon rich atmosphere will produce a 
stellar wind where the nucleation rate per unit mass is 
much higher than in case of normal abundance.
A lower gas to dust ratio than found in the 
ISM  has to be expected.
The stellar luminosity increased from about 3000 to 
about 8000 L$_\odot$. The 
dust formation rate ($\propto$ mass loss rate) is correlated 
about linearly to the stellar luminosity.
All the luminosities, provided here, are directly proportional 
to $D^{-2}$. The masses are proportional to $D^{-3}$.
With decreasing or increasing the free distance parameter $D$, the
derived mass would strongly change, too.

\vspace{-2mm}
\acknowledgements
\vspace{-1mm}
This  project  was  supported by the Austrian FWF  project P11675-AST.

\end{article}

\vspace{-1mm}
\begin{thebibliography}{}



\bibitem[1990]{Be90}
Bessel M.S. 1990, PASP 102, 1181

\bibitem[1994]{Bo94}
Borkowski K.J., Harrington J.P., Blair, W.P., Bregman J.D. 1994, ApJ 235, 722

\bibitem[1997]{Cl97}  
Clayton G.C., deMarco O. 1997, AJ 114, 2679


\bibitem[1993]{Do93}
Dominik C., Sedlmayr E., Gail H.P. 1993, A\&A 277, 578


\bibitem[1996]{Du96}  
D\"urbeck H.W., Benetti S. 1996, ApJ 468, L111

\bibitem[1997]{Du97}  
D\"urbeck H.W., Benetti S., Gautschy A., et al. 1997, AJ 114, 1657

\bibitem[2000]{Du00} 
D\"urbeck H.W., Liller W., Sterken C., et al. 2000, AJ 119, 2360




\bibitem[1990]{Ga90}
Gauger A., Sedlmayr E., Gail H.-P. 1990, A\&A 235, 345




\bibitem[1984]{Ib84}
Iben I. 1984, ApJ 277, 333

\bibitem[1983]{Ib83}
Iben I., Kaler J.B., Truran J.W., Renzini R. 1983, ApJ 264, 605




\bibitem[1999]{Kam99}  
Kamath U.S., Ashok N.M. 1999, MNRAS 302, 512

\bibitem[1998]{Ke98}
Kerber F., Gratl H., Kimeswenger S., Roth M. 1998, Ap\&SS 255, 279

\bibitem[1999]{Ke99}
Kerber F., Blommaert J.A.D.L., Groenewegen M.A.T., et al. 1999, A\&A 350, L27

\bibitem[2000]{Ke00}
Kerber F., Palsa R., K\"oppen J., Bl\"ocker T., Rosa M.R. 2000, The Messenger, 101, 27

\bibitem[1997]{Ki97}
Kimeswenger S., Gratl H., Kerber F., et al. 1997, IAUC 6608


\bibitem[1998]{Ki98} 
Kimeswenger S., Kerber F., Weinberger R. 1998, MNRAS 296, 614


\bibitem[1999a]{Kip99a} 
Kipper T. 1999a, IBVS 4707

\bibitem[1999b]{Kip99b} 
Kipper T. 1999b, Baltic Astronomy 8, 483



\bibitem[2000a]{Ko00a}
Koller J., Kimeswenger S. 2000, ASPConf. Ser. 196, 23

\bibitem[2000b]{Ko00b}
Koller J., Kimeswenger S. 2000, Ap\&SS, in press 

\bibitem[1996]{Na96} 
Nakano S., Benetti S., D\"urbeck H.W. 1996, IAUC 6322


\bibitem[1998]{Pa98}
Patzer A.B.C., Gauger A., Sedlmayr E. 1998, A\&A 337, 847

\bibitem[1999]{Pa99}
Patzer A.B.C., Helling Ch., Winters J.M., Sedlmayr E. 1998, Astron. Ges. Abstr. Ser. 15, P41




\end{thebibliography}
\end{document}